\documentclass[aps,twocolumn,prb,showpacs,eqsecnum]{revtex4}
\usepackage{amsmath,amssymb,latexsym}
\usepackage[dvips]{graphicx}

\begin{document}

\title{Phonon and plasmon excitation in inelastic electron tunneling
spectroscopy of graphite}

\author{L. Vitali, M. A. Schneider, and K. Kern}
\affiliation{Max-Planck-Institut f\"ur Festk\"orperforschung,
Heisenbergstr.1, D-70569 Stuttgart, Germany}
\author{L. Wirtz and A. Rubio}
\affiliation{Department of Material Physics, University of the
Basque Country, Centro Mixto CSIC-UPV, and Donostia International
Physics Center (DIPC), Po. Manuel de Lardizabal 4, E-20018 San
Sebasti\'an, Spain}

\date{\today}

\begin{abstract}
The inelastic electron tunneling spectrum (IETS)of highly oriented
pyrolitic graphite (HOPG) has been measured with scanning
tunneling spectroscopy (STS) at 6K. The observed spectral features
are in very good agreement with the vibrational density of states
(vDOS) of graphite calculated from first principles. We discuss
the enhancement of certain phonon modes by phonon-assisted
tunneling in STS based on the restrictions imposed by the
electronic structure of graphite. We also demonstrate for the
first time the local excitation of surface-plasmons in IETS which are
detected at an energy of 40 meV.
\end{abstract}

 \maketitle

Initially applied to measure inelastic tunneling currents across
metal-insulator-adsorbate-metal junctions \cite{wolf}, Inelastic
Electron Tunneling Spectroscopy (IETS) has in the recent past been
revived and its concept extended to Scanning Tunneling Microscopy
(STM-IETS) \cite{stipe}. Superimposed on the elastic tunneling
current, the inelastic tunneling process opens up alternative
tunneling channels increasing the total conductance at the onset
of its excitation. Very recently, elegant experiments (e.g.
\cite{stipe,kim,lee,pascnat}) have been performed in detecting
local inelastic electron energy-loss spectra of individual
molecule adsorbed on metallic substrates. The observed features in
the tunneling spectra are found to correspond to the vibrational
modes of the adsorbed species. Vibrations of single molecules as
large as C$_{60}$ have have been detected \cite{pascual}. Despite
the increasing number of publications of STM-IETS experiments on
single molecules adsorbed on metals, to our knowledge, only two
studies were reported on collective vibrational excitations of
surfaces \cite{smith,olejn}. In both these studies graphite was
used as a substrate and several spectral features remained
unexplained or were attributed to vibrational excitation in the
tungsten tip. However, these tip effects have not been
detected in many other STM-IETS experiments. In order to resolve
the pure inelastic response of a graphite surface, we have
performed STM-IETS experiments on a highly oriented pyrolitic
graphite crystal (HOPG) at cryogenic temperature. In our
experiments the whole vibrational spectrum of HOPG from the rigid
layer shear mode at $\approx$ 5meV to the optical modes at
$\approx$ 200 meV is clearly observed without being perturbed by
inelastic effects from the tip. The phonon modes of graphite can
be clearly identified by the direct comparison of experimental
data with first principles density function theory (DFT)
calculations. We report for the first time that also
surface-plasmon losses can also be observed directly with
STM-IETS.

Freshly cleaved HOPG has been introduced into the vacuum system
and shortly annealed at 500K. The experiments were performed at 6K
with a home built UHV-STM using a tungsten tip. The inelastic
tunneling spectrum of HOPG was measured over a topographically
clean and flat terrace with a lock-in technique. In this technique
a sinusoidal signal with amplitude $\Delta V$ is superimposed to
the applied sample bias.  In IETS the second harmonic component of
the modulated signal, which is proportional to the second
derivative vs. bias of the tunneling current, $d^2I/dV^2$, is
recorded. In this way the inelastic loss features in the
conductance are transformed into peaks at the onset of the loss
modes for positive sample bias and into minima for negative sample
bias, respectively. A typical example for HOPG is shown in fig.1. 
\begin{figure}[htpb]
 \centering
   \includegraphics[draft=false,keepaspectratio=true,clip,%
                   width=1.0\linewidth]%
                   {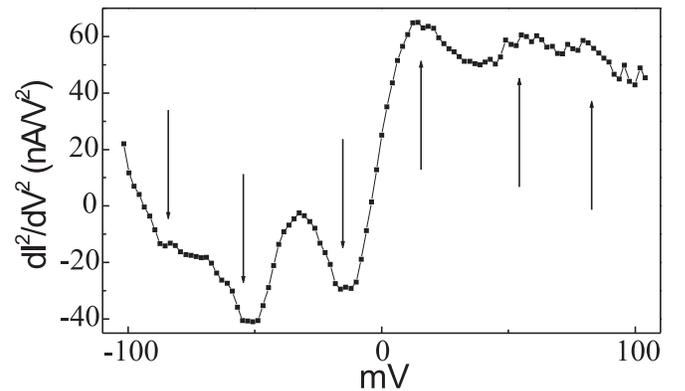}
\caption{Inelastic electron tunneling spectrum of HOPG measured by
STM. The phonon features (arrows) in the $d^2I/dV^2$ are symmetrically located 
with respect to the zero bias position.} \label{samplefig}
\end{figure}

We obseve phonon features that are symmetrically located with respect
to zero bias as expected for the inelastic tunneling process.
Possibly due to a particular tip-end configuration teh phonon on the negative bias side
are more pronounced. In Fig. 2a we present a high-resolution spectrum of the $d^2I/dV^2$
signal of HOPG taken at negative bias. The data was obtained with a bias modulation of $\Delta V =
10$~meV RMS at a frequency of 2.7kHz and inverted to render
the phonon excitations as peaks to facilitate the comparison to
theory. Since the electron density of states of HOPG is
flat in the energy range close to the Fermi level, the features in
the spectrum in Fig. 1 are exclusively due to inelastic processes.
\begin{figure}[htpb]
 \centering
   \includegraphics[draft=false,keepaspectratio=true,clip,%
                   width=1.0\linewidth]%
                   {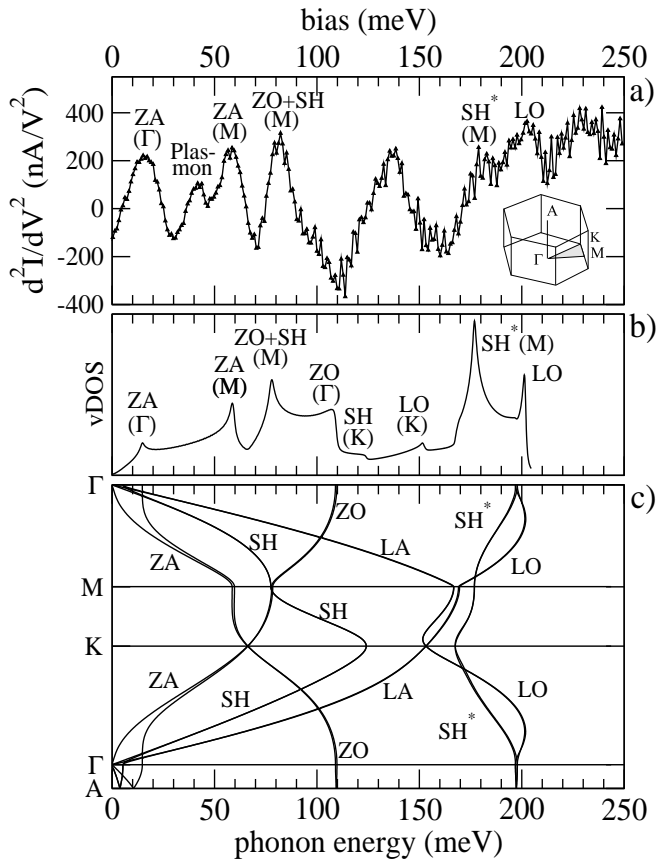}
\caption{ a) Inelastic electron tunneling spectrum of HOPG
measured by STM. Inelastic excitations show up as peaks in the
$d^2I/dV^2$ signal. b) vibrational density of states calculated by
DFT-LDA. c) Phonon dispersion relation of graphite. The phonon
branches are specified as following: out of plane acoustic (ZA),
acoustic shear (SH), longitudinal acoustic (LA), out of plane
optical (ZO), optical shear (SH$^*$), and longitudinal optical
(LO). The inset shows the first Brillouin zone of graphite with
the high symmetry points. A detailed description of the observed
phonon modes is given in the text. Furthermore, a plasmon
excitations is identified at 40 meV.} \label{samplefig}
\end{figure}

Along with the measured inelastic spectrum, we present in Fig. 2b
and c the calculated vibrational density of states (vDOS) and the
phonon dispersion relation of graphite along the A-$\Gamma$-M-K
directions. Graphite is a layered structure with strong carbon
bonds in the plane and weak Van der Waals-like forces between the
planes. As a result of this anisotropy, the phonon spectrum of
graphite covers a very wide range. We have performed first
principles calculations using density-functional theory (DFT) in
the local density approximation (LDA)~\cite{calcdetails}. Aim of
this calculation is to resolve some of the deviations in the vDOS
calculated by the different recent force-constant parametrizations
for graphite \cite{jishi,gruneis,aizawa,sieben} and to have a
solid foundation for the assignment of the measured peaks. The
first principles calculation has been shown
\cite{kresse,pavone,dubay} to yield very close agreement with
phonon dispersion relations measured by high resolution electron
energy loss spectroscopy (HREELS) along the $\Gamma$-M direction
\cite{aizawa} and along the $\Gamma$-K direction \cite{sieben}.
Only for the acoustic shear-mode (SH) it was observed \cite{dubay}
that at the M-point, the HREELS data of Ref.\onlinecite{aizawa} yields
an energy which is higher by 25 meV than the {\it ab-initio}
value. A similar underestimation of of the shear-mode frequency
might occur as well at the K-point where a mode with energy 134
meV was observed by double-resonance Raman scattering
\cite{kawashima}. Our calculated vDOS is also at variance with the
HREELS data at M \cite{aizawa} but is in very good agreement with
the vDOS measured by neutron scattering on a powdered sample of
graphite \cite{rols} where a separate SH(M)-feature was not
observed. However, since the SH K-point mode contributes only
weakly to the calculated DOS the comparison to the experimental
vDOS in Ref. \onlinecite{rols} cannot answer the question whether the
mode comes out too soft in the calculations.

\begin{table}
\begin{tabular}{|c|c|c|c|c|} \hline
Phonon  & Symmetry  & STM-IETS & Other experiments &
DFT-LDA \\%
branch & point &   (meV)           & (meV) & (meV)\\
\hline
ZA     & $\Gamma$ & 16  & 16 \cite{nicklow} & 15 \\
ZA     & M        & 58  & 57 \cite{aizawa}  & 59 \\
ZO     & M        & 81  & 81 \cite{aizawa}  & 78 \\
SH     & M        & 100 (weak) & 100 \cite{aizawa}  & 77 \\
ZO     & $\Gamma$ & 111 (dip)  & 108 \cite{sieben} & 109  \\
LA     & K        & 137 & 134 \cite{kawashima} & 124 \\
SH$^*$ & M        & 180 & 172 \cite{aizawa} & 177 \\
LO     & $\Gamma$ & 200 & 198-205 \cite{aizawa} & 197-202 \\
\hline
\end{tabular}
\caption{ Observed features in the STM-IET spectrum of HOPG in
comparison with vibrational modes observed in experiment (Neutron
scattering \cite{nicklow}, HREELS \cite{aizawa,sieben} and
double-resonant Raman scattering \cite{kawashima}) and with first
principles DFT-LDA calculations.} \label{scaletable}
\end{table}

The observed vibrational spectral features, their comparison with
values obtained with other experimental techniques and their
assignment to phonon modes at the various points of the surface
Brillouin zone (SBZ) are summarized in Table 1. Most features
coincide with peaks in the calculated vDOS which arise from the
flat dispersion relation around the high-symmetry points. In
particular, the higher peaks due to phonon modes around the
M-point are clearly mapped. The first peak at 16~meV is in good
agreement with the calculated energy of the out of plane acoustic
(ZA) mode at $\Gamma$ where two neighboring planes are oscillating
out-of-phase. While in the vDOS and in neutron scattering
\cite{rols} this peak is only weak, in the IETS experiment it is
stronlgy pronounced which can be explained by the direction of
oscillation which is orthogonal to the surface plane. The weak
shoulder at 6~meV corresponds to the low frequency horizontal mode
at $\Gamma$ where two neighboring planes are oscillating
out-of-phase parallel to each other. The prominent feature at
40~meV is not due to a vibrational feature but is assigned to a
plasmon excitation (see below). In Ref.\onlinecite{smith}, a phonon
feature at 39~meV was observed which was associated with a
double-phonon excitation in the surface (16~meV) and in the
tungsten tip (23~meV). However, in our experiment the tip-phonon
peak is clearly not present and all the observed features can
therefore be assigned to surface excitation. The vibrational
feature observed at 58~meV can be assigned to the ZA at the
M-point. The optical out-of-plane (ZO) mode and the shear acoustic
mode SH (also called transverse acoustic mode) display a saddle in
the dispersion relation around the M-point adding up to a very
high peak in the vDOS at 78~meV. This explains the peak at 81 meV
in the measurement. However, as noted above, HREELS data
\cite{aizawa} suggests the energy of the SH mode at M to be about
100~meV. This gives rise to the alternative explanation that the
peak at 81~meV is only due to the ZO mode while the shoulder at
100~meV may be a signal of the SH mode at the M-point. The ZO mode
at $\Gamma$ which gives rise to a peak at 108 meV in the
theoretical vDOS seems to be missing in the spectrum. Instead, the
spectrum reveals actually a dip ($dI^2/dV^2 < 0$) at this energy
which will be discussed below. At 124 and 150 meV two
low-intensity phonon modes, the SH, and the longitudinal acoustic
(LA) modes at the K-point are predicted, respectively.
Experimentally we have observed a large asymmetric inelastic
feature at about 137~meV. This disagreement between experiment and
theory might point to an underestimation of the SH mode also at
the K point. If, as could be the case for the SH mode at M
\cite{dubay}, the real energy is higher by 10 to 20~meV, the peak
at 137~meV can be assigned to the SH mode at K which corresponds
to the above mentioned observation of a K-point phonon at 134~meV
\cite{kawashima} by double resonance Raman spectroscopy. In the
highest energy range of the spectrum two loss-features protrude
the background signal at 180 and 200~meV, respectively. The first
is related to the optical shear mode (SH$^*$, also called
transverse optical mode) around the M-point. The neighboring
feature at 200~meV corresponds to the longitudinal optical (LO)
branch which displays a strong overbending in the direction
$\Gamma$-M and $\Gamma$-K giving rise to a peak in the vDOS.

In general, the measured IET spectrum follows very closely the
vibrational density of states with two notable exceptions which
will be discussed in the following. These exceptions are related
to the question if all phonons can be excited or if the geometry
of the tunneling system leads to a restriction to phonons in a
certain region of the Brillouin zone, as it is the case in
phonon-assisted electronic tunneling through narrow {\it p-n}
junctions of semiconductors. In e.g. Ge, the valence-band maximum
is centered at $\Gamma$ while the conduction-band minimum is
located at the $L$ point. This allows only vibrational modes at L
to be excited by a phonon-assisted tunneling process \cite{payne}.
The electronic structure of graphite is also quite unique being a
semimetal with a band crossing only at the K-point of the SBZ.
Therefore the elastic current is carried by electrons that tunnel
with a relative large $k_\parallel =$~K (1.7~\AA$^{-1}$). In an STM
geometry the expansion of the initial state, i.e., the tip wave
function in terms of surface plane-waves \cite{tersoff}, leads to
a parallel momemtum distribution centered at $k_\parallel = 0$.
Consequently the electrons of the elastic current have a low
transmission probability. Electrons with any other $k_\parallel$
can only tunnel by exchange of finite crystal momentum with a
phonon. The phonon-assisted tunneling is relevant to all the
dominant features observed besides the $\Gamma$ phonons at 16 and
200~meV. An enhanced tunneling contribution should especially
occur for electrons with initial parallel momentum $k_\parallel =
0$ that tunnel to the Fermi level at the K-point by exciting (or
absorbing) a K-point phonon. This explains why the peak at 137 meV
which is probably due to a K-point phonon excitation shows up so
dominantly in the spectrum even though the vDOS of states only
displays weak features for K-point phonons.

The second deviation of the measured spectrum from the vDOS of
graphite is the dip at 111 meV where the vDOS displays a peak due
to the ZO phonon at $\Gamma$. The theory for inelastic tunneling
spectroscopy \cite{persson,lorente} distinguishes between an
inelastic contribution (real phonon emission or absorption) and an
elastic contribution (virtual phonon emission with subsequent
reabsorption or phonon absorption with subsequent re-emission). In
some cases, the elastic contribution leads to a decrease of the
tunneling conductance (dI/dV) due to back-scattering of the
electron~\cite{persson}. This effect has been measured recently as
dips in the $dI^2/dV^2$ curve of O$_2$ on a silver surface
\cite{hahn}.  We believe that the dip at 111 meV is due to a
similar elastic contribution at the onset of vDOS singularity
corresponding to the ZO($\Gamma$) phonon branch. In the light of
the above discussion of phonon-assisted tunneling on graphite, it
is also not surprising that this effect is observed for phonons at
$\Gamma$. The only matching pair for a virtual phonon scattering
process are electrons initially at $k_\parallel = K$ and phonons
at $\Gamma$. All other combinations are forbidden due to the
unavailability of intermediate electronic states. Furthermore, the
steady reduction of $-d^2I/dV^2$ from the ZO (M) peak at 81~meV,
in contrast to the vDOS, shows that symmetry of the phonon
displacement (the ZO branch) should play an important role in the
supression of the IET current~\cite{lorente2}. We note that one of
the degenerate modes of the ZO branch at $\Gamma$ is the only
$\Gamma$ phonon mode that has $A_{2u}$ symmetry.

In addition to the predicted phonon modes, the measured spectrum
contains an energy loss feature at 40 meV, which cannot be
explained by phonon excitations. Other than the previous features,
this loss feature is of purely electronic origin and corresponds
to an out-of-plane oscillation of the electrons. A tunneling
electron traversing a vacuum-metal interface has a significant
probability of interacting with collective electron oscillations.
The overlapping of the $\pi$ and $\pi^*$ band at the K point of
the SBZ yields a low density of free carriers (i.e. electrons and
holes) near the Fermi level and allows low-energy plasmon
excitation. A strong temperature dependence of the excitation
energy of this plasmon mode was observed in HREELS \cite{jensen}.
The excitation energy shifts from 63 meV to 45 meV when going from
room temperature to 150 K, respectively. This effect has been
attributed to the low density of states at the Fermi level in
conjunction with the thermal excitations of carriers. Thermal
excitations from the valence into the conduction band can
radically alter the electron population at the Fermi level and
yields the strong dependence of surface plasmon energy and line
width on temperature. The temperature dependence of this
excitation energy was also calculated in Ref. \onlinecite{jensen}.
Within the random phase approximation the plasmon energy
excitation was predicted to be lowered to 36 meV at 0K in rather
good agreement with our experimental value. The low-energy plasmon
was also observed in transmission electron energy loss
spectroscopy, which also detected a second plasmon loss, assigned
to an excitation polarized parallel to the layer at 128 meV
\cite{geiger}. The shoulder observed around 125 meV might be
assigned to this plasmon excitation.

In conclusion, the IET spectrum of HOPG has been measured by low
temperature scanning tunneling spectroscopy and is compared to an
ab-initio calculation of the vibrational density of states in
graphite. The observed vibrational loss features can be identified
with peaks in the vDOS which are due to the flat dispersion of the
phonon branches around the high-symmetry points of the Brillouin
zone. Additionally, a prominent plasmon feature at 40meV has been
identified in the inelastic tunneling current measured with STM.
In the past, the local capability of STM-IETS has been proven to
be a valuable tool to access the vibrational properties of
adsorbates at surfaces. In this work we have shown that STM-IETS
resolves the vibrational and collective electronic excitations of
graphite with high accuracy. This is an important step for the
application of the method to confined geometries and individual
carbon nanostructures.
\acknowledgments 
L.W. and A.R. thank support from the EC research training network 
COMELCAN (HPRN-CT-2000-00128), Spanish MCyT(MAT2001-0946) and UPV/EHU
(9/UPV 00206.215-13639/2001).



\begin{thebibliography}{99}
\bibitem{calcdetails}
The calculations have been performed with the code ABINIT
\cite{abinit}. The force constants are calculated within
linear-response theory \cite{baroni}. The wave-functions are
expanded in plane waves with an energy cutoff at 80 Ry. Core
electrons are described by Trouiller-Martins pseudopotentials. The
1st Brillouin zone is sampled by a 16x16x6 Monkhorst-Pack grid.
The phonon frequencies are converged to within 4 cm$^{-1}$ with
respect to variation of the energy cutoff and variation of k-point
sampling. The dynamical-matrix, which is the Fourier-transform of
the real-space force constants, is calculated on a 14x14x6
Monkhorst-Pack grid in the reciprocal space of the phonon-wave
vector $q$. From this, the dynamical matrix at any $q$ is obtained
by interpolation.
\end{thebibliography}
\end{document}